\documentclass[prl,twocolumn,preprintnumbers,amsmath,amssymb,showkeys]{revtex4-1}

\usepackage{graphicx}
\usepackage{dcolumn}
\usepackage{bm}
\usepackage{amsmath,amssymb}
\usepackage{mathrsfs}

\begin{document}

\title{Demonstration of a microfabricated surface electrode ion trap}

\author{D.~Stick}
 \email{dlstick@sandia.gov}
 \author{K.~M.~Fortier}
\author{R.~Haltli}
\author{C.~Highstrete}
\author{D.~L.~Moehring}
\author{C.~Tigges}
\author{M.~G.~Blain}

\affiliation{Sandia National Laboratories, Albuquerque, New Mexico 87185, USA}

\date{\today}

\begin{abstract}
In this paper we present the design, modeling, and experimental testing of surface electrode ion traps fabricated in a heterostructure configuration comprising a silicon substrate, silicon dioxide insulators, and aluminum electrodes. This linear trap has a geometry with symmetric RF leads, two interior DC electrodes, and 40 individual lateral DC electrodes.  Plasma enhanced chemical vapor deposition (PECVD) was used to grow silicon dioxide pillars to electrically separate overhung aluminum electrodes from an aluminum ground plane.  In addition to fabrication, we report techniques for modeling the control voltage solutions and the successful demonstration of trapping and shuttling ions in two identically constructed traps. 
\end{abstract}


\maketitle

\section{Introduction}
\label{sec:intro}
Individually trapped ions are a leading candidate for quantum information processing \cite{blatt:2008,wineland:2009}, as most of the DiVincenzo requirements have already been substantially realized \cite{divincenzo:2000}.  Of these requirements, the current limiting factor is whether trapped ions constitute a ``scalable physical system'', due to the difficulty of creating large trapping structures capable of independently controlling tens or hundreds of ions.  There has been an increasing emphasis on creating scalable architectures \cite{wineland:1998,kielpinski:2002,rowe:2002,madsen:2004,stick:2006,hensinger:2006,reichle:2006,brownnut:2006,leibfried:2007,amini:2008,blakestad:2009}, with surface traps being an especially promising approach due to their compatibility with standard fabrication techniques like photolithography, via technology, wire bonding, and metal evaporation \cite{chiaverini:2005,britton:2006,seidelin:2007,brown:2007,allcock:2009,britton:2009,leibrandt:2009,hellwig:2009}.  Most importantly, junctions and backside loading holes can be incorporated in a surface geometry, the latter of which we discuss in this paper.

\section{Fabrication}
\label{sec:fab}
Surface electrode traps have been previously demonstrated by a number of groups \cite{seidelin:2007,brown:2007,allcock:2009,leibrandt:2009} using a variety of techniques (e.g. gold on quartz, printed circuit board, aluminum on silicon oxide). The traps reported here are similar in spatial scale and geometry to these traps; however, particular emphasis is placed on the design principle of minimizing the line of sight access to the ion from exposed dielectrics, thereby reducing the impact of stray electric charges.  Numerous observations have been made of shifting trapping potentials over long time scales (seconds) due to changes in the location and magnitude of stray charges \cite{harlander:2010}.  

\begin{figure}
\resizebox{0.45\textwidth}{!}{%
  \includegraphics{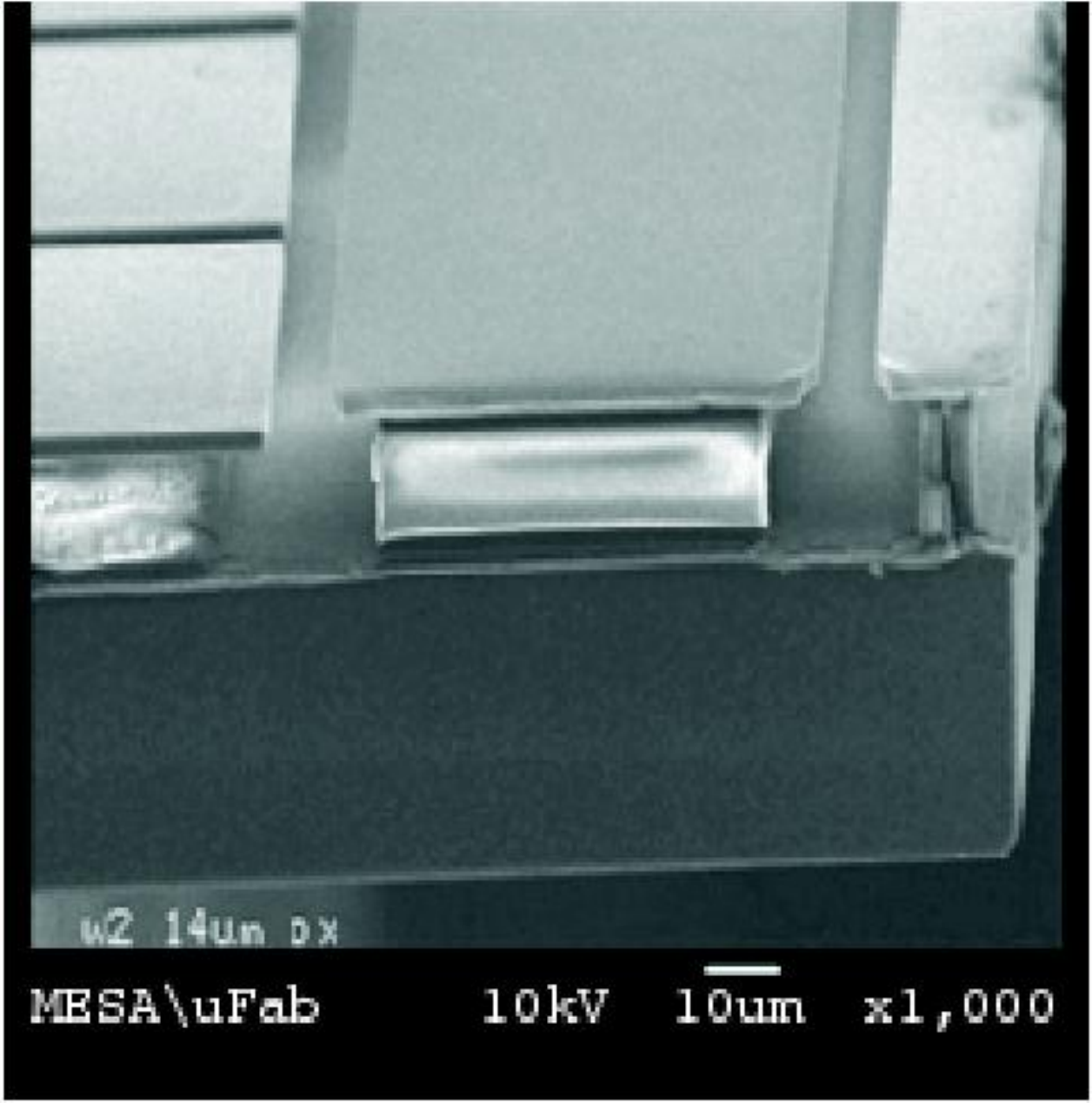}}
\caption{SEM image showing the 5 micron overhang from the supporting oxide pillar and the 7 micron gap between neighboring electrodes.}
\label{fig:etchstop}       
\end{figure}

\begin{figure}
\resizebox{0.45\textwidth}{!}{%
  \includegraphics{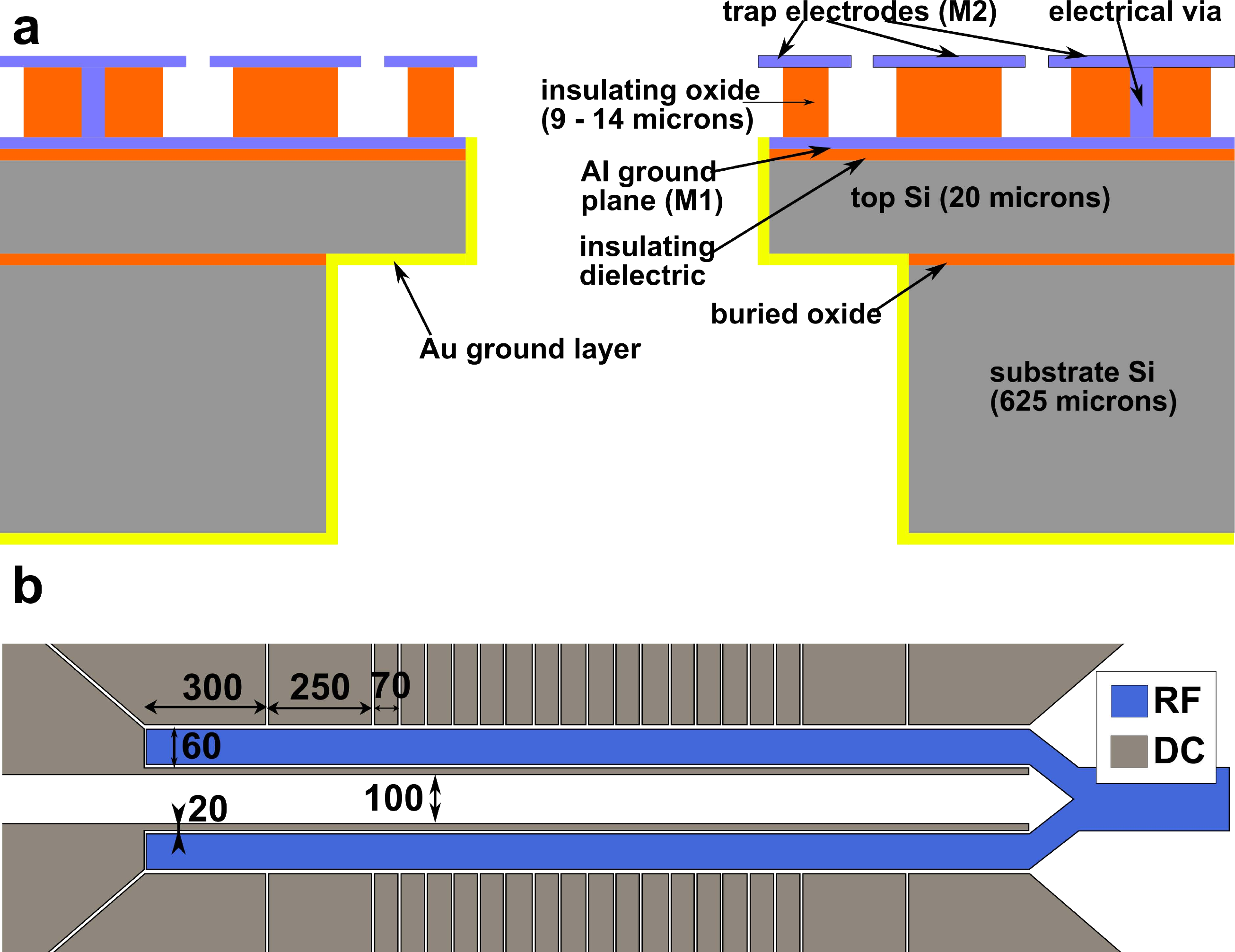}}
\caption{Cross-sectional (a) and overhead (b) schematic of the ion trap.}
\label{fig:schematic}       
\end{figure}

\begin{figure}
\resizebox{0.45\textwidth}{!}{%
  \includegraphics{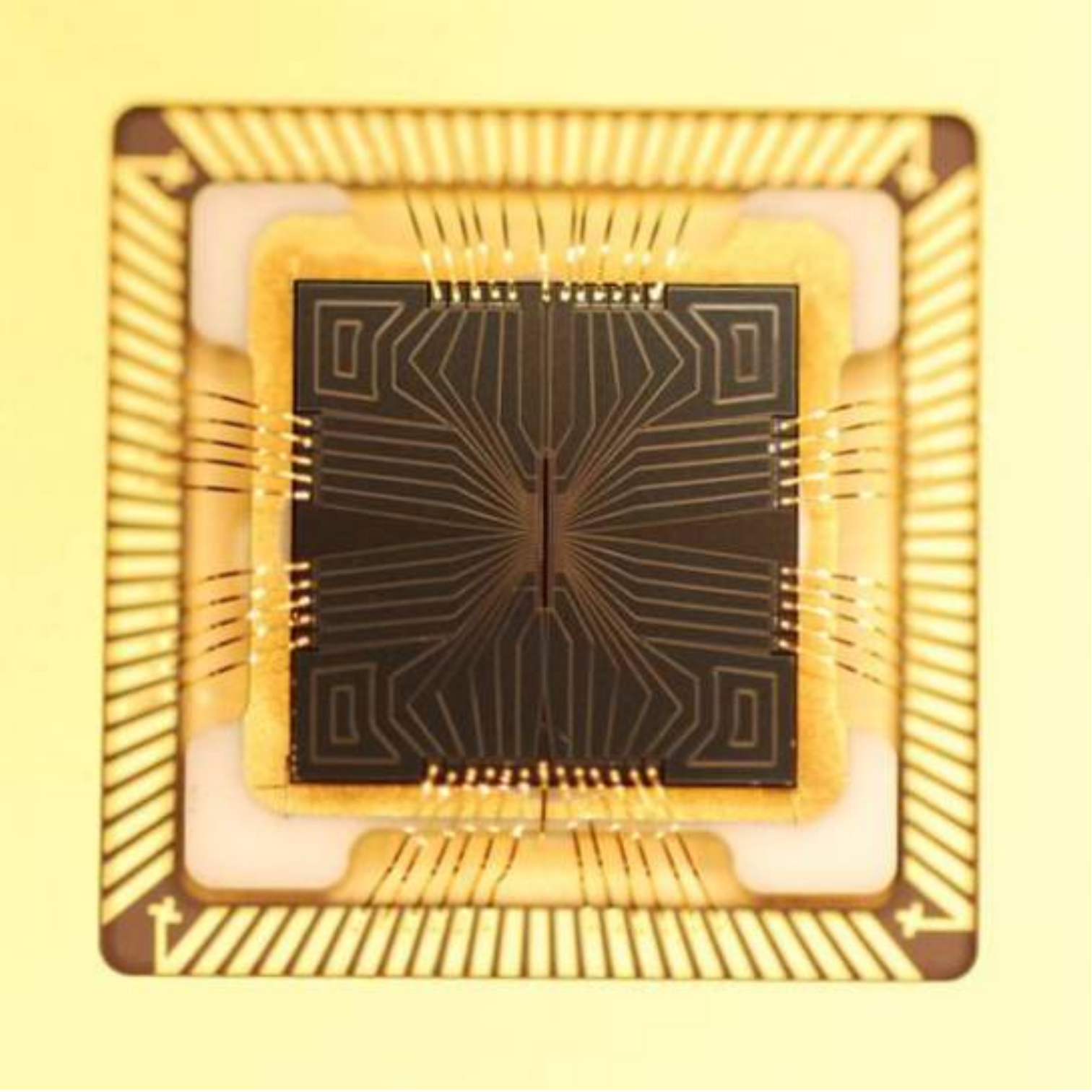}}
\caption{Electrode layout.}
\label{fig:layout}       
\end{figure}

To realize this design principle, the top metal layer of these traps (comprising electrodes, their leads, and outside grounded regions) overhang their supporting oxide pillars by 5 $\mu$m. The oxide pillars are grown through multiple layers of plasma enhanced chemical vapor deposition, and are between 9 and 14 $\mu$m thick (the traps reported here have 9 $\mu$m pillars).  The overhang distance is a controllable value achieved by using vertical etch stops around the pillars (Figure~\ref{fig:etchstop}). The overhang allows for vertical deposition of metal on top of the aluminum electrode layer without shorting DC control or RF electrodes. The lateral separation between electrically isolated top metal layers (such as between neighboring electrodes) is set to be 7 microns, and the lateral dimensions of the electrodes can be arbitrarily determined (see Figures~\ref{fig:schematic} and \ref{fig:layout} for specific dimensions).  A hole through the Si substrate of the trap chip runs the entire length of the trapping region to allow for loading of ions from the backside of the trap (preventing shorting of the trap electrodes by the atoms, which can occur when loading ions from the side).   DC rails inside the RF rails allow for additional principle axis rotation and compensation. The back side of the chip is evaporated with gold at a small off-normal angle to coat the exposed vertical edges of the silicon substrate and the platform which supports the electrodes.  This prevents charge buildup by pinning the backside of the chip to ground (Figure~\ref{fig:schematic}a).  


Once fabricated, the chip is mounted in a 100 pin ceramic pin grid array (CPGA) package from Kyocera
[21].  A conductive cyanate ester adhesive (Johnson Matthey Electronics JM7000) is used to attach the chip to a 1.5 mm thick ceramic spacer, which holds the surface of the trap above the surface of the package. The package and chip back side surfaces are gold coated to electrically connect and ground the back of the package, the chip, the epoxy, and the vertical silicon sidewalls (Figure~\ref{fig:schematic}). 

Gold ribbons (12.5 $\mu$m thick by 75 $\mu$m wide) are wedge bonded to each electrode at I/O pads located on an electrical plane (M1) 9 $\mu$m beneath the trapping electrode plane (M2) and pulled taut to the package bond pads to minimize their projection above the plane of the trap surface.  The lower metal plane, M1, serves primarily as a ground plane to prevent RF coupling into the lossy silicon substrate.  Each trap is electrically tested for shorts between any trap electrodes (DC, RF, and grounds, including the ground plane).  For results using both traps reported here, no shorts below 100 M$\Omega$ were observed. The RF capacity of the trap is measured by applying an RF drive (30 MHz at 2 Watts of power) to the trap through a resonator with a Q of 150. The total capacitance from RF to ground (trap and CPGA package) is 7 pF.  The control electrodes are capacitively grounded outside of the vacuum chamber, and low pass filters (3 kHz or 200 kHz, depending on whether shuttling tests are being performed) are used to attenuate electronic noise.

\section{Modeling}
\label{sec:modeling}
A custom boundary element modeling approach was used to find the charge solution for all 42 control electrodes and the RF electrode.  The area of each element was chosen such that each element had the same charge; this corresponds to larger elements at points farther away from the electrode in question, and serves to standardize the error per element.  The RF null was then determined by finding the minimum pseudo-potential along the linear trapping region.  Control solutions were generated in order to maximize a weighted figure of merit that includes secular frequency, trap depth, and principal axis rotation, such that the electric field at a particular position along the RF null is zero.  The solution was verified using a flight simulator which determined the ion's motion according to the electric field at its position (including both the control solution and an oversampled RF drive).  The ion's flight was verified for thousands of times longer than the period of the RF drive voltage, and the secular frequencies determined by Fourier transforming the ion's motion.

The simulations were experimentally validated by applying a voltage to each control electrode such that the ion moved a fixed distance of 2 $\mu$m axially.  This was repeated many times and an image was take for each offset.  The ion's location was precisely determined using a Gaussian fit of the image, and the simulations corresponded to the experimental measurements of the ion's motion to within the error of the position measurement (10\%).

\section{Trap Performance}
\label{sec:perform}
The ion is observed to be trapped 80 $\mu$m above the top electrode layer, consistent with simulations.  The uncooled ion lifetime in the first trap was 3-5 minutes, and was observed to be sensitive to DAC cable shielding (lifetimes dropped to $\sim$10 s without twisted pair shielding).  The cooled ion lifetime remained on the order of several hours throughout this period.  Two ion traps were tested side-by-side in separate UHV chambers, using identical control voltage sets for storage and shuttling.  The traps were operated at a wide range of RF drive frequencies, although trapping multiple Ca$^+$ ions was easier with a $43$~MHz RF drive frequency compared to a $27$~MHz drive frequency.  

The secular frequencies for a given voltage set (typical frequencies are 1 MHz axial and 4 MHz radial) were measured by observing driven motion for resonant tickling voltages \cite{jefferts:1995} and by measuring the separation between two trapped ions \cite{wineland:1998}, and were consistent over time. The observed drift of the ion was measured to be $\leq .5 \mu$m (axially) over a 250 s period of observation, and depended on the power of the UV laser beams (Doppler and photoionization) and the extent to which they struck the surface. To prevent charge buildup on the inside of the imaging viewport, a wire mesh (88\% open) was attached inside the vacuum chamber, 1.5 mm above the surface of the trap.  Through simulations it was determined that this would have minimal impact on the trapping potential.  

\begin{figure}
\resizebox{0.45\textwidth}{!}{%
  \includegraphics{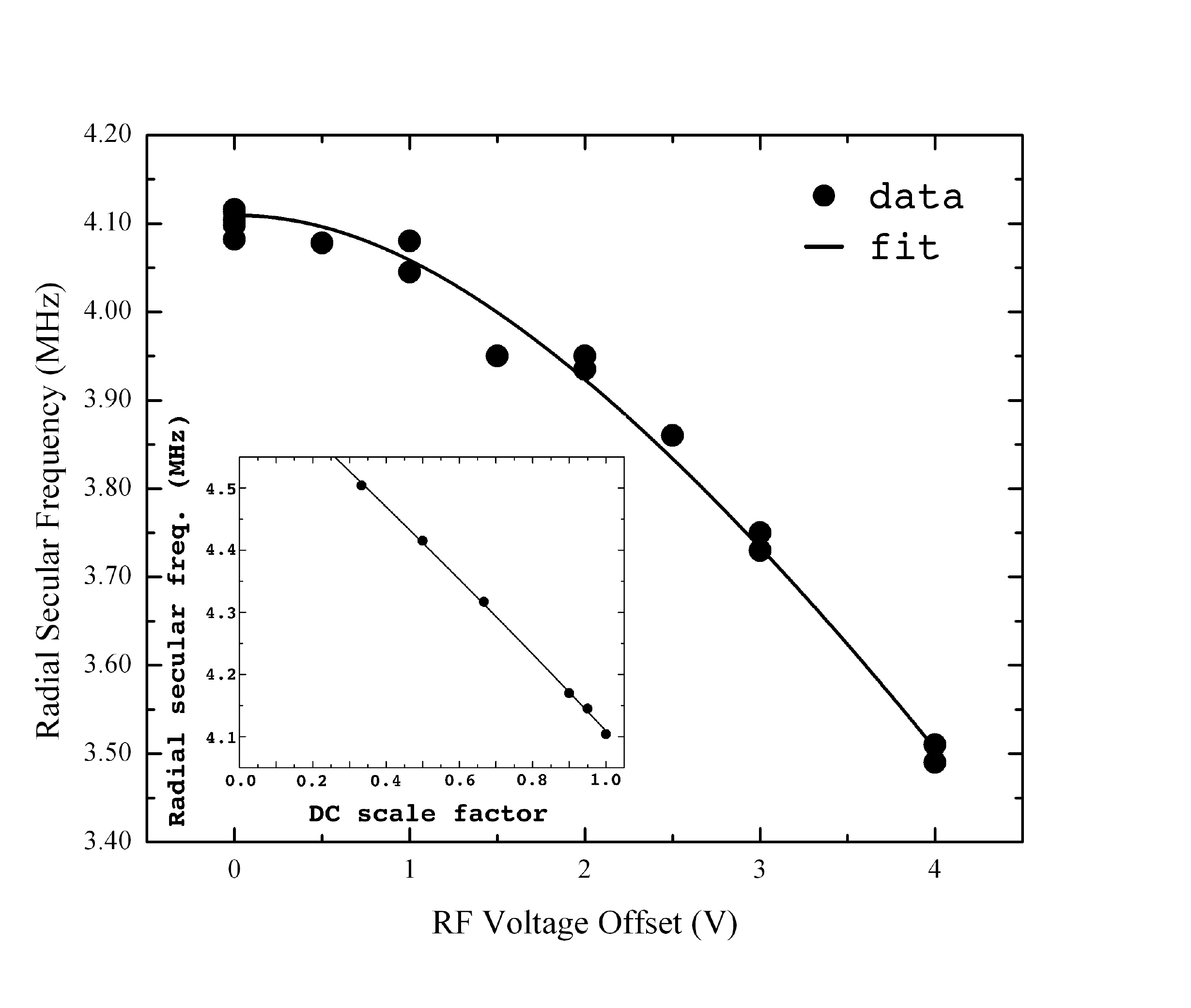}}
\caption{The secular frequency was measured as a function of varying voltage offsets $U_r$ to the RF electrode.  Fitting this curve allowed us to determine the amplitude of the RF voltage and the principal axis rotation, for a particular DC voltage set and RF power applied.  For this particular case, the RF amplitude was determined to be 140 V (amplitude) and the principal axis rotation was 39 degrees from vertical.}
\label{fig:secular}       
\end{figure}

The RF voltage is delivered through a cavity resonator with a Q $\approx100$, and an amplitude between 50 V and 200 V.  By measuring the change in radial and axial secular frequencies when scaling a particular DC voltage set at a fixed RF voltage, the geometric potential factors were determined for the control electrodes.  An applied DC offset to the RF electrodes changes both the radial secular frequencies and the rotation of the principal axes.  The RF voltage amplitude and principal axis rotation can be determined to within a few percent by fitting these frequencies to a numerical model (Figure~\ref{fig:secular}), and agreed with electrostatic simulations to 10\%. 

Motional control was demonstrated by shuttling a single ion over half the length of the trapping structure (10 electrodes, 770 microns) for $10^6$ times without loss.  This is a total travel distance of just over 1.5~km, and was performed at a maximum average velocity of .77 m/s.  Ion chains have also been split into two parts and recombined.  Future work on this trap will include measurements of the induced heating of the ion for these shuttling and splitting operations.

\section{Conclusion}
\label{sec:conclusion}

For trapped ions to be a suitable platform for quantum computing, a scalable-in-principle technique for trap fabrication has to be demonstrated.  The surface geometry is the most amenable to microfabrication, but it poses challenges related to the low trap depth and difficulty in making a working shuttling junction. Our demonstration of a microfabricated surface electrode trap addresses this first challenge, and given its consistency of fabrication and trap performance, can be used to create more sophisticated structures with similarly repeatable performance.

\end{document}